\documentclass[usenatbib,usegraphicx]{mn2e}

\usepackage{psfig}
\usepackage{amsmath}

\def\apj{ApJ}
\def\aap{A\&A}
\def\aj{AJ}
\def\prd{Phys. Rev. D}
\def\ssr{Space Sci. Rev.}
\def\mnras{MNRAS}
\def\nat{Nature}

\def\cm{\textrm{cm}}
\def\erg{\textrm{erg}}
\def\kpc{\textrm{kpc}}
\def\pc{\textrm{pc}}
\def\Mpc{\textrm{Mpc}}

\def\Kelv{\textrm{K}}
\def\kms{\textrm{km}~\textrm{s}^{-1}}
\def\meV{\textrm{meV}}
\def\eV{\textrm{eV}}

\def\GeV{\textrm{GeV}}

\def\yr{\textrm{yr}}

\def\Myr{\textrm{Myr}}
\def\Gyr{\textrm{Gyr}}
\def\Tyr{\textrm{Tyr}}
\def\nGauss{\textrm{nG}}
\def\muGauss{\mu\textrm{G}}
\def\Gauss{\textrm{G}}
\def\Msun{\textrm{M}_{\sun}}
\newcommand{\mean}[1]{\ensuremath{\langle #1 \rangle}}

\title[Cosmic Rays in Intergalactic Medium]{The Ultimate Fate of Cosmic Rays from Galaxies and their Role in the Intergalactic Medium}
\author[Lacki]{Brian C. Lacki\\Institute for Advanced Study, Einstein Drive, Princeton, NJ 08540, USA, brianlacki@ias.edu}

\begin{document}

\maketitle

\begin{abstract}
The majority of cosmic rays (CRs) generated by star-forming galaxies escape them and enter the intergalactic medium (IGM).  Galactic wind termination shocks might also accelerate CRs.  I show that the mean pressure of these CRs can reach to within an order of magnitude of the mean Lyman-$\alpha$ forest thermal pressure.  At $z \ga 1$, their pressure may have even been dominant.  I also demonstrate that, whichever IGM phase the CRs reside in, they contribute significantly to its pressure if its temperature is $\sim 10^4\ \Kelv$, as long as pionic and Coulomb losses are negligible.  Where CRs end up depends on the structure and strength of intergalactic magnetic fields.  I argue that CRs end up at least 30 kpc from their progenitor galaxies.  CRs may self-confine in the IGM to the sound speed, generating $\ga 10^{-13}\ \Gauss$ magnetic fields.  These considerations imply the existence and importance of a nonthermal IGM.  
\end{abstract}


\section{Introduction}
Most of the baryonic matter in the Universe is not in stars or galaxies, but in the gas between the galaxies, the intergalactic medium (IGM).  Far from being a uniform background gas, the IGM has a rich structure with several phases spanning a wide range of physical conditions.  The scaled densities $\delta \equiv \rho / (\Omega_b \rho_{\rm crit})$ within the IGM vary by orders of magnitude.\footnote{$\rho_{\rm crit}$ is the critical density $3 H_0^2 / (8 \pi G)$.}  At present, about one-third of the baryonic mass is in the Lyman-$\alpha$ (Ly$\alpha$) forest, a volume-filling $10^4$ -- $10^5\ \Kelv$ photoionized phase with $\delta \la 10$ \citep{Bi97}.  Before $z \approx 1$, this phase contained most of the baryonic mass of the Universe \citep{Dave99}.  As some of this gas collapses into smaller structures, it is shock heated to form the Warm Hot Intergalactic Medium (WHIM), a moderately dense ($\delta \sim 10 - 100$), $10^5$ -- $10^7\ \Kelv$ phase that presently contains $\sim 1/2$ of the baryonic mass \citep{Cen99}.  Finally, some of the material has collapsed even further into `condensed' structures: these include the galaxies themselves, their multiphase gaseous haloes, and hot intracluster media.  The circumgalactic gas results from the interaction of galaxies and the IGM, through accretion on to galaxies and expulsion by winds, and is itself complex, with cool dense and warm rarefied phases \citep*[e.g.,][]{Chen01}.  

But is this picture of the IGM complete -- an interplay of thermal gas, galaxies, and gravity?   Just as in galaxies, the thermal gas may be complemented by pervasive nonthermal fields: the cosmic rays (CRs) and magnetic fields.  Detections of radio synchrotron emission within galaxy clusters prove that both CRs and magnetic fields exist there, and much work centres on how they are generated and interact with the intracluster gas (e.g., \citealt*{Subramanian06}; \citealt{Ferrari08}).  Radio emission is also detected from galaxy filaments, suggesting similar processes at play (\citealt{Brown09}; see also \citealt{Ryu08}).  

Evidence for nonthermal processes in the rest of the IGM is scant, but there are theoretical reasons to expect they exist.  A nonthermal IGM can take the form of CRs accelerated by structure formation shocks (\citealt{Loeb00}; \citealt*{Keshet04}), injected by dark matter annihilation throughout the Universe \citep*{Mapelli06}, or escaping from active galactic nuclei jets \citep{Vecchio13}; pervasive intergalactic magnetic fields (IGMFs) seeded by early galaxies, plasma processes in the IGM, or even primordially in the Big Bang (\citealt{Widrow02}; \citealt{Schlickeiser12-B}; \citealt*{Yoon14}); and low frequency nonthermal radio waves that heat the IGM when absorbed \citep{Lacki10-SubMHz}.  Whether electromagnetic TeV $\gamma$-ray cascades heat the IGM nonthermally through plasma instabilities (\citealt*{Broderick12}; \citealt*{Chang12}) is the topic of intense recent debate \citep{Schlickeiser12-Beam,Miniati13}, although recent simulations suggest this mechanism is fairly ineffective \citep{Sironi14}.

A guaranteed source of nonthermal energy in the IGM is the CRs accelerated by star-forming galaxies (SFGs).  Most of the CR energy is in GeV protons.  They diffuse out of SFGs, losing only a few percent of their energy through pion production and ionization in Milky Way-like galaxies (\citealt{Strong10}, hereafter S10). Starbursts and high-$z$ normal galaxies have more gas, and likely heavier losses.  Yet, $\gamma$-ray observations of the starbursts M82 and NGC 253 indicate that $\sim 20 - 40\%$ of their CR power is lost to pion production, with the majority ($\sim 60 - 80\%$) escaping, most likely in a starburst wind \citep{Lacki11-Obs,Abramowski12,Ackermann12}.  The termination shocks of these winds may also accelerate CRs \citep{Jokipii85,Jokipii87}.  Thus, a large reservoir of CRs builds up in the IGM over the Gyrs of cosmic star formation.  

This population has been largely ignored.  \citet{Nath93} and \citet*{Samui05} studied how CRs from SFGs could have reionized the IGM, or heated it at high redshift.  But the pionic energy loss time for CRs with kinetic energies $\ga \GeV$ is $\sim 200\ \delta^{-1}\ \Tyr$, so energy injection is negligible in the rarefied IGM (\citealt{Mannheim94,Schlickeiser02}, hereafter S02).  \citet{Miniati11} proposed that CRs from the first SFGs seeded the IGMF.  A few other estimates of the intergalactic CR spectrum, with widely disparate assumptions, are briefly given in \citet{Dar05} and \citet{Lipari05}. 

CRs can nevertheless couple dynamically to the IGM and its weak magnetic fields.  I show here that the mean intergalactic energy density of CRs from SFGs is plausibly within a factor of a few of the Ly$\alpha$ forest pressure (Section~\ref{sec:IGMPressure}).  I also show that CRs are likely to reach the IGM (Section~\ref{sec:WhereCRs}).  I assume $\Omega_{\Lambda} = 0.75$, $\Omega_m = 0.25$, $H_0 = 70\ \kms\ \Mpc^{-1}$ for the cosmology.  For the baryonic density, I use $\Omega_b = 0.045$, a hydrogen mass fraction $X_H = 0.75$, a helium mass fraction $X_{\rm He} = 0.25$, and mean molecular weight $\mu = 0.6$ (complete ionization) for a mean IGM comoving number density of $\mean{n_{\rm IGM}} = 4.1\ \times 10^{-7}\ \cm^{-3}$.\footnote{The mean comoving electron density is then $\mean{n_e} = (X_H + X_{\rm He} / 2) \mu \mean{n_{\rm IGM}} = 2.2 \times 10^{-7}\ \cm^{-3}$.}  All given densities are comoving unless otherwise stated.

\section{The Pressure of Intergalactic Cosmic Rays}
\label{sec:IGMPressure}

CR nuclei experience mainly adiabatic losses once they leave a galaxy.  CRs in the IGM lose momentum as the Universe expands.  CRs in a galactic wind lose momentum adiabatically as the wind expands, sacrificing their energy to push the wind outwards \citep*{Volk96}.  Galactic winds are common in SFGs with high specific star-formation rate, both high-$z$ main sequence galaxies and compact starbursts (\citealt*{Heckman90}; \citealt{Steidel10}).  Those winds are powered at least in part by the thermalization of supernova (SN) mechanical energy.  SFGs without SN-launched winds may still have more rarefied winds originating from their haloes and powered by CRs, as their streaming excites plasma waves that push halo plasma out (\citealt*{Breitschwerdt91}, hereafter B91).  The existence of this kind of CR-driven halo wind is unproven -- CRs may even simply diffuse out without energy losses (as assumed in \textsc{galprop} models; S10).  

The equation for the comoving IGM particle momentum spectrum at redshift $z$ is
\begin{equation}
\label{eqn:dNdlnp}
\frac{dN}{d\ln p}(p, z) = \int_z^{\infty} \frac{d^2 Q}{d\ln p_{\rm inj} dt} (p_{\rm inj}, z_{\rm inj}) \frac{d\ln p_{\rm inj}}{d\ln p} \frac{dt}{dz_{\rm inj}} dz_{\rm inj},
\end{equation}
Particles are accelerated to momenta $p_{\rm inj}$ at a rate $d^2 Q/d\ln p_{\rm inj} dt$ at redshift $z_{\rm inj}$.  I relate $p_{\rm inj}$ to $p$ as $p = \epsilon_{\rm adv} p_{\rm inj} (1 + z) / (1 + z_{\rm inj})$, where $\epsilon_{\rm adv}$ accounts for adiabatic losses within a galactic wind.  For purely adiabatic losses, $d\ln p_{\rm inj} / d\ln p = 1$.  I consider Coulomb and pionic losses later, as they are negligible if $\delta \sim 1$.

The injection spectrum is modelled as a power law in momentum ($d^2 Q/(dp_{\rm inj}\ dt) = C p_{\rm inj}^{-\Gamma}$), with spectral index $\Gamma = 2.2$.  The normalization of the spectrum is set by a volumetric energy injection rate:
\begin{equation}
\label{eqn:CRNorm}
\dot{\varepsilon}_{\rm CR} = \int_{{\rm GeV}/c}^{{\rm PeV}/c} C K p_{\rm inj}^{-\Gamma} dp_{\rm inj},
\end{equation}
where $K$ is kinetic energy.  I calculate $\dot{\varepsilon}_{\rm CR}$ by scaling to the comoving cosmic star-formation rate density $\rho_{\rm SFR}$ \citep{Hopkins06}.  Star formation creates young, massive stars that accelerate CRs in their winds or SNe.  The star-formation rate is directly related to the SN rate as $\Gamma_{\rm SN} = 0.0084\ \yr^{-1} [{\rm SFR} / (\Msun\ \yr^{-1})]$ for the `Salpeter A' initial mass function used by \citet{Hopkins06}.  Multiwavelength models of SFGs indicate that each SN accelerates $\sim 10^{50} \eta_{0.1}\ \erg$ of CRs (that is, $10 \eta_{0.1}\%$ of the $10^{51}\ \erg$ released in mechanical energy; \citealt{deCeaDelPozo09}; S10; \citealt{Lacki11-Obs,YoastHull13}). CRs may also be accelerated by galactic wind termination shocks, where the winds are powered by SNe and contain some fraction $\epsilon_{\rm SN}$ of the original SN mechanical energy.  So I have
\begin{equation}
\label{eqn:EnergyInjection}
\dot{\varepsilon}_{\rm CR} = 0.018\ \meV\ \Gyr^{-1}\ \cm^{-3}\ \left(\frac{\rho_{\rm SFR} \times \eta_{0.1} \epsilon_{\rm SN}}{1\ \Msun\ \yr^{-1}\ \Mpc^{-3}}\right).
\end{equation}

Relativistic CRs advected out from their host galaxy in a wind of density $\rho$ lose momentum as $\rho^{1/3}$, so the $\epsilon_{\rm adv}$ factor is simply the cube root of the ratio of the final and initial wind density.\footnote{The momentum scaling is due to the isentropic nature of these processes.  Phase space volume is conserved, so as the real space volume goes as $\rho^{-1}$, the momentum space volume scales with $\rho$, and the momentum in one direction scales as $\rho^{1/3}$.}   For example, a hypothetical B91 wind from the Milky Way initially occupies the disc with radius 10 kpc and the CRs move with its Alfv\'en speed of $10\ \kms$.  If the wind outflows spherically and reaches 1 Mpc with a speed of $300\ \kms$ (B91), then $\epsilon_{\rm adv} = [(4 \pi (1\ \Mpc)^2 (300\ \kms)) / (2 \pi (10\ \kpc)^2 (10\ \kms))]^{-1/3} \approx 0.012$.  Adiabatic cooling is much greater within winds from compact starburst regions, because the density contrast from the starburst to the IGM is huge.

The adiabatic energy scaling of CRs ranges from $\rho^{2/3}$ for non-relativistic ($\la \GeV$) CRs to $\rho^{1/3}$ for relativistic CRs ($\ga \GeV$).  CRs with initial Lorentz factors $\la \epsilon_{\rm adv}^{-1} (1 + z)$ are thus severely affected by adiabatic losses and do not contribute much to the IGM CR energy density.  Yet since the injection spectrum is close to $p^{-2}$, much of the original energy is in very high energy CRs that always are relativistic.  For a $p^{-2.2}$ injection spectrum, the fraction of kinetic energy in always-relativistic CRs is roughly $[\epsilon_{\rm adv}^{-1} (1 + z)]^{-0.2}$; this cutoff factor is $0.4 (1 + z)^{-0.2}$ for $\epsilon_{\rm adv} = 0.01$.  Even for a $p^{-2.4}$ spectrum, the cutoff factor $0.16 (1 + z)^{-0.2}$ is $\ga 0.1$.  Hence, the increased losses of non-relativistic CRs only have a moderate effect on the final CR energy density.  

Although a wind could drain most of its CR energy by the time it reaches the IGM, the wind itself eventually stops at a termination shock when its ram pressure equals the surrounding IGM's pressure.  The termination shock itself may accelerate CRs, converting $\ga 10\%$ ($\eta_{0.1} \ga 1$) of the wind's kinetic energy back into CRs and releasing them into the IGM \citep{Jokipii85}.  However, the efficiency of CR acceleration in collisionless shocks depends on the conditions within the shock \citep[e.g.,][]{Caprioli14}.  These conditions are especially unknown in galactic wind shocks, as is whether CR acceleration actually takes place there. The B91 wind's kinetic energy is powered by the CRs initially present in the galaxy, which themselves contain only $10\%$ of the original SN power so that $\epsilon_{\rm SN} = 0.1$.  Powerful starburst winds may carry most of the SN mechanical power, implying $\epsilon_{\rm SN} = 1$.  Reacceleration can also occur at shocks within inhomogeneous winds \citep{Dorfi12}.

\begin{figure}
\centerline{\includegraphics[width=8cm]{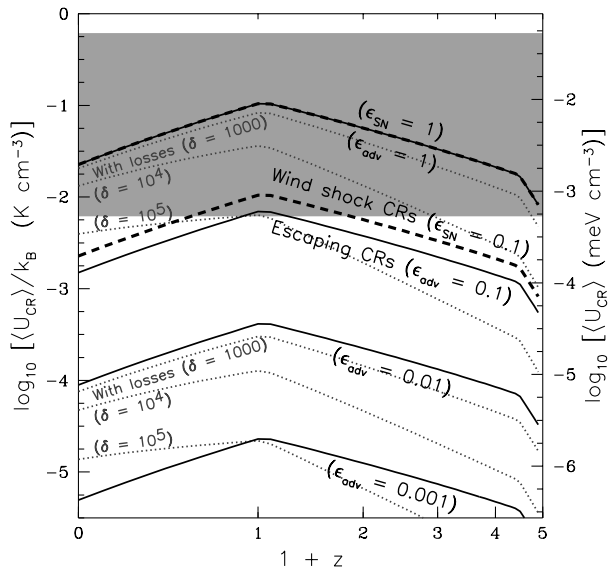}}
\caption{Predicted comoving mean cosmic CR energy density.  The dashed lines are for CRs accelerated at wind termination shocks, while solid lines represent the original escaping population of CRs after experiencing adiabatic losses.  Grey dotted lines show models including Coulomb and pionic losses with $\delta = 1000, 10^4, 10^5$ for wind termination shocks (upper three lines) or adiabatically cooled CRs ($\epsilon_{\rm adv} = 0.01$; lower three lines).  The grey shaded band is the expected mean IGM thermal pressure for $T = 10^4$ K -- $10^6$ K.  All plotted models assume that $\eta_{0.1} = 1$.  \label{fig:UCREvolution}}
\end{figure}

I find that with $\epsilon_{\rm adv} = 1$, the comoving CR energy density $\mean{U_{\rm CR}} / k_B$ is $0.02\ \Kelv\ \cm^{-3}$ at present, and it peaked at $z = 1.0$ at $0.1\ \Kelv\ \cm^{-3}$ (Figure~\ref{fig:UCREvolution}).  About half of the present energy density is in CRs accelerated at $z < 0.4$ (in the past 4 Gyr).  For termination shocks CRs with $\epsilon_{\rm SN} = 0.1$, $\mean{U_{\rm CR}} / k_B$ is ten times smaller.  For comparison, the expected comoving thermal IGM energy density is $0.0075 T_4\ \Kelv\ \cm^{-3}$ for $T = 10^4 T_4\ \Kelv$ (grey band in Figure~\ref{fig:UCREvolution}).  We see that, if $\epsilon_{\rm adv} = 1$, intergalactic CRs have an energy density greater than that of the Ly$\alpha$ forest.  For $\epsilon_{\rm shock} = 0.1$, CRs and Ly$\alpha$ forest thermal energy are in equipartition at $z \approx 1$.

Pionic and Coulomb losses matter on cosmological time-scales for GeV CRs if $\delta \ga 1000$.  Galactic winds remain fairly dense even $\sim 100\ \kpc$ from their launch sites, with $\delta \ga 1000$ in cool gas at $z \approx 0$ \citep[e.g.,][]{Narayanan08,Werk14}.  Thus, I add these losses and consider cases when $\delta = 1000, 10^4$, and $10^5$.  I integrate all momentum losses over redshift to relate $p$ and $p_{\rm inj}$.  I assume that all CRs are protons.  The Coulomb loss rate is $\dot{p}_{C} = 3.1 \times 10^{-7} \delta \mean{n_e} (1 + z)^3 \beta_{\rm CR}^{-2} (\eV\ c^{-1}\ \sec^{-1})$ and the pionic loss rate is $\dot{p}_{\pi} = p \delta \mean{n_e} (1 + z)^3 / (50\ \Myr)$ for $p \ge 0.78\ \GeV/c$ (\citealt{Mannheim94}; S02).\footnote{$\beta_{\rm CR} = pc / (K + mc^2)$ is the CR speed in units of $c$.}

The grey dotted lines in Figure~\ref{fig:UCREvolution} represent the volume-mean CR energy density if CRs are trapped in high-$\delta$ regions.  If $\delta = 10^5$, the CR energy density is reduced by a factor 70 (6) at $z = 5$ (0), whereas the reduction is only 1.7 (1.2) if $\delta = 1000$.

\subsection{CR importance in overdense IGM phases}
CRs might not actually fill the entire volume of the Universe, but instead may be confined to one IGM phase.  Although these phases could be very overdense, the CR energy density is likewise greater since the same CR power is then being squeezed into a smaller confinement volume.

What is the ratio of CR energy density to thermal energy density in the phase the CRs end up in?  If a fraction $f_i$ of the CR energy occupies an IGM phase $i$, and if $i$ has a cosmic filling factor $\phi_i$, then the CR energy density within $i$ is $U_{\rm CR} = f_i \phi_i^{-1} \mean{U_{\rm CR}}$.  The gas density in $i$ is limited by mass conservation to $n_i = (\Omega_i / \Omega_b) \phi_i^{-1} \mean{n_{\rm IGM}}$, where $\Omega_i$ is the the baryonic mass fraction in $i$.  The thermal energy density within that phase is $U_{\rm IGM} = (3/2) n_i k_B T_i$, giving
\begin{equation}
\frac{U_{\rm CR}}{U_{\rm IGM}} = f_i \frac{\Omega_b}{\Omega_i} \frac{2 \mean{U_{\rm CR}}}{3 \mean{n_{\rm IGM}} k_B T_i}.
\end{equation}

The filling factors cancel out: if CRs end up mostly in one phase of the IGM, then all that matters are its mass fraction and its temperature (neglecting Coulomb and pionic losses).  $U_{\rm CR}$ exceeds the thermal energy density as long as $T_i < (2/3) f_i (\Omega_b / \Omega_i) [\mean{U_{\rm CR}} / (\mean{n_{\rm IGM}} k_B)]$ and $\delta \la 1000$.

At $z = 0$, the CRs dominate the energy density of their final IGM phase if $T_i < 3.6 \times 10^4 \Kelv (\Omega_b / \Omega_i) f_i \Phi$, where $\Phi \approx 0.1$ -- 1 is the ratio of the actual $\mean{U_{\rm CR}}$ and its value for $\epsilon_{\rm adv} = 1$, $\epsilon_{\rm SN} = 1$, and $\delta \ll 1000$.  About half of the $z = 0$ IGM mass has a temperature $\sim 10^4\ \Kelv$, including the Ly$\alpha$ forest and condensed haloes, indicating a major CR contribution to their pressure is possible.  However, half of the IGM mass is WHIM with $10^5 - 10^7\ \Kelv$ temperatures; if the CRs end up in WHIM, their pressure is insignificant.

At higher redshift, CRs are even more important because $\rho_{\rm SFR}$ was much greater.  Furthermore, most of the gas was in the uncollapsed $10^4\ \Kelv$ Ly$\alpha$ forest.  I find that CRs dominate the pressure of their host phase at $z = 1$ if $T_i < 1.6 \times 10^5\ \Kelv (\Omega_b / \Omega_i) f_i \Phi$, a condition that applied to most of the IGM's mass.

\subsection{Do CRs couple to the IGM?}

In order for the CR pressure to affect the dynamics of the IGM, the CRs must interact with it.  This happens whenever the CRs' paths bend.  When the IGM exerts a magnetic force on the CRs, the CRs exert a force back on the IGM.

The trajectories of CRs may bend for several reasons.  CRs gyrate around field lines if there is a coherent magnetic field, but this acceleration has a time average of zero.  CRs can self-confine and scatter off plasma waves in the IGM, diffusing parallel or perpendicular to magnetic field lines.

Yet even without these waves, CRs move along magnetic field lines.  If the lines bend, then the CRs' paths also bend, and the CRs exert a force on the IGM.  CRs are deflected within some region of light-crossing time $t$ if the IGMFs have coherence length $\lambda_B$ smaller than the region but larger than $r_L \equiv pc/(eB)$.  The required magnetic field strength is then $B \ga p / (e t) = 4 \times 10^{-22}\ \Gauss\ p_{\GeV} t_{10}^{-1}$, where $t_{10} = t / (10\ \Gyr)$ and $p = p_{\GeV} \GeV / c$.

If $\lambda_B$ is smaller than $r_L$ and $ct$, the CRs can still be deflected after passing through many IGMF domains.  Each domain exerts an impulse $\delta p = e B \lambda_B / c$, where $\delta p \ll p$.  A CR's trajectory bends only when the sum of all the impulses is $\ga p$.  This condition is met when $B > p c / (e \sqrt{c t \lambda_B}) = 2 \times 10^{-20}\ \Gauss\ p_{\GeV} / \sqrt{t_{10} (\lambda_B / \Mpc)}$.  Several plausible mechanisms exist for generating coherent IGMFs this strong (e.g., \citealt{Widrow02,Miniati11}), and blazar $\gamma$-ray spectra indicate their presence \citep{Dermer11}.

Thus, CRs are likely to provide significant pressure support in some IGM phase at $z \ga 1$.  If the CR pressure is completely homogeneous, then it cannot exert net forces.  CRs that are trapped in collapsing IGM regions, though, are adiabatically heated, leading to pressure gradients.  This may alter the equation of state of the IGM, for example, but the effects of the CRs remain unexplored.

\section{Where are the Cosmic Rays?}
\label{sec:WhereCRs}

\subsection{Do CRs really even escape their galaxies?}

Observations of quasar absorption lines indicate that winds from high-$z$ SFGs transport material out to $\ga 100\ \kpc$ \citep{Heckman90}.  Theory likewise suggests that even Milky Way-like galaxies host CR-driven winds in their haloes (B91).  Ultimately, a wind with asymptotic speed $v_{\infty}$ and kinetic luminosity $\dot{E}$ should flow out to the termination shock radius $R_s = \sqrt{\dot{E} / (4 \pi v_{\infty} P_{\rm IGM})} \approx 1.1\ \Mpc\ [\dot{E} / (10^{40}\ \erg\ \sec^{-1})]^{1/2}\ [v_{\infty} / (1000\ \kms)]^{-1/2}\linebreak[0]\ [P_{\rm IGM} / k_B / (0.005\ \Kelv\ \cm^{-3})]^{-1/2}$.  It takes roughly 1 (10) Gyr for a $1000\ (100)\ \kms$ wind to traverse 1 Mpc.  Thus, galactic winds can transport energy deep into the IGM.

But suppose CRs are not advected out but simply diffuse from their host galaxies (as in \textsc{galprop} models)?  CR diffusion results in a net flow out of the Galaxy only if the energy density outside is less than that inside.  Since CRs \emph{do} actually diffuse out of the Galaxy, the CR energy density in the distant Galactic halo is less than that within the disc.  For a steady CR luminosity, we therefore have $(t_{\rm MW} / V_{\rm out}) < (t_{\rm in} / V_{\rm in})$.  The Milky Way has been forming stars for $t_{\rm MW} \approx 10\ \Gyr$, whereas the time that CRs stay within the Galactic disc is only about $t_{\rm in} \approx 30\ \Myr$ \citep{Connell98}.  The confinement volume of the Galactic disc is $V_{\rm in} \ga 350\ \kpc^3$ (S02).  We can set limits on $V_{\rm out}$, the confinement volume of CRs that have `escaped' the Galaxy.  Supposing that the CRs fill a sphere with radius $R_{\rm out}$, I conservatively find:
\begin{equation}
R_{\rm out} \ga \left(\frac{3}{4\pi} V_{\rm in} \frac{t_{\rm MW}}{t_{\rm in}}\right)^{1/3} = 30\ \kpc \left(\frac{V_{\rm in}}{350\ \kpc^3}\right)^{1/3}.
\end{equation}
If CRs are not advected away, they \emph{must} diffuse far beyond the Galactic disc, well into the circumgalactic gas.

\subsection{How far in the IGM do they go?}
The propagation of CRs depends on the IGMF strength and structure \citep{Adams97}.  Very little is known about the IGMF, other than that it less than $\la 1\ \muGauss$ for all coherence lengths and $\la 1\ \nGauss$ on large scales \citep{Neronov10}.  The most interesting constraints come from $\gamma$-ray observations.  The lack of GeV cascade emission from blazars suggests that pair $e^{\pm}$ generated by TeV $\gamma$-rays on their way to Earth are deflected out of the sightline by IGMFs.  This sets a lower limit of $10^{-18}\ \Gauss$ \citep{Dermer11}, but the applicability of the limits is disputed \citep{Broderick12}.

Supposing that CRs do reach the large-scale IGM, Bohm diffusion represents the slowest possible propagation.  The mean free path is $\lambda_{\rm CR} = r_L \approx 1.1\ \Mpc\ p_{\GeV} B_{-18}^{-1}$, where $B = 10^{-18}\ B_{-18}\ \Gauss$. The distance CRs diffuse is then $s_{\rm Bohm}\ \approx\ \sqrt{{\beta_{\rm CR}} c t {\lambda_{\rm CR}}}\ { = 58\ \Mpc\ \sqrt{\beta_{\rm CR} t_{10} p_{\GeV} / B_{-18}}}$.  Note that $s_{\rm Bohm} < 1\ \kpc$ for $B \ga 10\ \nGauss$. Bohm diffusion is only an extreme limit, requiring big enough $\lambda_B$; CRs probably diffuse much farther than $s_{\rm Bohm}$.

The maximum distance that CRs propagate into the IGM could be set by plasma waves that they excite while they stream.  Within galaxies, where the magnetic energy density is comparable to thermal pressure, CRs self-confine themselves to speeds less than the Alfv\'en speed \citep{Kulsrud69}.  In the classical version of CR self-confinement, the magnetic fluctuations from CR streaming are much smaller than the mean field.  However, it is thought that some non-linear version of the instability can amplify magnetic fluctuations until they are larger than the mean field to confine CRs if necessary \citep{Lucek00}.  

The sound speed is much greater than the Alfv\'en speed in the IGM, though.  In these conditions, the streaming speed of CRs is probably of order the sound speed $c_s = 15\ \kms \sqrt{T_4}$ \citep*{Holman79}.  This weak self-confinement limits the distance that intergalactic CRs stream to $s_{\rm confine} = c_s t = 0.15\ \Mpc\ t_{10} \sqrt{T_4}$.  Note that in hot phases like the WHIM, the CRs can stream out fastest.  For $T = 10^6\ \Kelv$, $s_{\rm confine}\ =\ 2\ \Mpc$, bigger than the typical size of a WHIM structure. 

Even if wave generation slows down CRs, this is very interesting in itself: it means that CRs excite IGMFs where they reside.  The CR mean free path is at least $r_L$.  If their average streaming speed over a time $t_{\rm stream}$ is $\la c_s$, then the magnetic field fluctuation strength must be at least ${\beta_{\rm CR} p c^2} / (e c_s^2 t_{\rm stream})$, or $B \ga 10^{-13}\ \Gauss\ p_{\GeV} T_4^{-1} [t_{\rm stream} / (10\ \Gyr)]^{-1}$ on some $\lambda_B$ larger than $10\ \pc\ p_{\GeV} (B/10^{-13}\ \Gauss)^{-1}$.  Any magnetic fluctuations excited by CRs are no greater than $\sim \sqrt{8 \pi U_{\rm CR}}$ on energetic grounds; for $z = 0$, this is $\la 9\ f_i \phi_i\ \nGauss$ for $\epsilon_{\rm adv} = 1$ (escaping CRs) or $\epsilon_{\rm SN} = 1$ (wind shock CRs).

\section{Conclusions}
The injection of CRs by SFGs into the IGM is a long-neglected kind of feedback.  The energy in CRs accelerated by SFGs over the Universe's history is comparable to the thermal energy in the Ly$\alpha$ forest.  Although the CRs may experience strong adiabatic losses as they are advected away from SFGs by winds, some of that energy may be recovered in CRs at the winds' termination shocks.  The calculated pressure of intergalactic CRs is then still within an order of magnitude of that of the Ly$\alpha$ forest.  If the CRs are trapped in a denser phase, their own density is also necessarily higher, meaning that these CRs are important if they end up in gas with $T \approx 10^4\ \Kelv$ and $\delta \la 1000$.  CRs may have been especially important at $z \ga 1$, when the star-formation rate density was greatest and most of the IGM was cool.

If standard theories of CR streaming apply to the IGM, the CRs affect its magnetic structure.  As they self-confine to roughly the sound speed, they excite magnetic fluctuations with $B \ga 10^{-13}\ \Gauss$.  Even so, they can penetrate through up to 100 kpc of $10^4\ \Kelv$ gas and 1 Mpc of $10^6\ \Kelv$ gas.  

SFGs are the not the only source of CRs in the IGM.  Active galactic nuclei and structure formation accelerate CRs too.  Rather, CRs from SFGs demonstrate the need to consider nonthermal processes in the IGM.  The nonthermal fields may prove to be just as important to the IGM as in the interstellar medium.

\section*{Acknowledgements}
I was supported by a Jansky Fellowship from the National Radio Astronomy Observatory when I first wrote this paper.  The National Radio Astronomy Observatory is operated by Associated Universities, Inc., under cooperative agreement with the National Science Foundation.  I also acknowledge support from the Institute for Advanced Study.  I thank the referee for their comments.

\end{document}